# Highly Relevant Routing Recommendation Systems for Handling Few Data Using MDL Principle and Embedded Relevance Boosting Factors

Diyah Puspitaningrum
University of Bengkulu
Bengkulu, Indonesia
diyahpuspitaningrum@gmail.com

I.S.W.B. Prasetya
Utrecht University
Utrecht, the Netherlands
S.W.B.Prasetya@uu.nl

P.A. Wicaksono
University of Bengkulu
Bengkulu, Indonesia
P.A.Wicaksono@gmail.com

## ABSTRACT

A route recommendation system can provide better recommendation if it also takes collected user reviews into account, e.g. places that generally get positive reviews may be preferred. However, to classify sentiment, many classification algorithms existing today suffer in handling small data items such as short written reviews. In this paper we propose a model for a strongly relevant route recommendation system that is based on an MDL-based (Minimum Description Length) sentiment classification and show that such a system is capable of handling small data items (short user reviews). Another highlight of the model is the inclusion of a set of boosting factors in the relevance calculation to improve the relevance in any recommendation system that implements the model.

## CCS CONCEPTS

• **Retrieval tasks and goals** → **Clustering and classification**; *Sentiment analysis*; Recommender systems • **Evaluation of retrieval results** → Relevance assessment;

## KEYWORDS

MDL classification, relevance scoring, boosting factor

## 1 INTRODUCTION

A *route recommendation system* is a system that helps users to find nearest places of interest. The system allows the user to type in a query to find a place, e.g. a restaurant. The system locates all the places within some preset distance that match the query, and then presents them in a ranked list, usually ordered by their routing distance. The user then selects one of them, and the system will then displays the corresponding route to the selected place. Nowadays, people also write reviews on places they visited. These comments are often available for public. Exploiting them can improve the relevance of the routes recommended by such a system, e.g. by preferring places with generally positive reviews. To be able to do this, the system must first be able to analyze the 'sentiments' of the reviews, e.g. whether a given comment indicates an excited user, or otherwise a bored one. However, one of main obstacles in applying sentiment analysis on user reviews is that these documents are *small* (most user reviews consist only of one or two sentences), and hence the information they individually convey is very limited as well.

One way to infer the sentiment that a piece of information conveys is by classifying the information into some domain of sentiments. So, it can be seen as a classification problem. Previously, the work by Sigurbjornsson and van Zwol (2008)[13] proposed to handle small data items by using pairwise tag co-occurences. The approach is both accurate and fast, but it needs larger sets of co-occuring tags to do the tagging; *tagging* here means assigning keywords or terms to a piece of information about an object. The work by Menezes et al. (2010)[9] used association rules instead and showed that exploiting association rules with more than two tags can improve accuracy. The association rule is employed to a database of sets of tags collection. Using a pattern selection method which is based on a compression technique such as Minimum Description Length (MDL) can further improve the accuracy [7]. A *pattern* here refers to a set of tags/terms, as the representation of documents viz. visitor reviews, that usually go together. A data structure such as *code table* can be used to store patterns' relative frequencies [15] and turns out to be able to accurately summarize a database. Alternatively, using a classification technique based on association rules mining, one can use MDL to mine very frequent patterns [8] in a sentiment database. This method [8] assigns more associated tags (two or more tags) and is slightly faster due to the use of code tables acting as summary databases to classify a data item.

In general, existing classifications techniques are either based on rules induction or based on association rules. A NaiveBayes (NB) technique [4] is frequently used for text classification because it is easy to implement but the drawback is that it tends to be slower and more complex because we must a priori specify the setting of model parameters, i.e. the prior probability. Even then, computing a posterior may be extremely difficult and this often leads to computationally infeasible situation. Another well-known technique is SVM [2] that has good performance but suffers in careful set up of key parameters. SVMs are also very sensitive to the problem that they are handling, so we end up needing different setups for different problems. An MDL-based classification technique is an NB-like algorithm, but it takes advantages of the availability of the distribution of sample data, encoded in a so-called 'code table'. If the database is an i.i.d. sample of some underlying data distribution, using an optimal code table of this distribution will well compress any arbitrary sample of records/transactions from the database. The MDL technique is expected to mine and select item-sets that are very characteristic for the data.

In our work, we are interested in the research question of whether a highly relevant route recommendation system can be built on top of an MDL-based sentiment analysis. This paper presents a model of such a system and an evaluation of the performance of an implementation of the model. In addition to an MDL back-end the model also incorporates a scoring function a la [6] that has been boosted with factors that include sentiment. We measured the performance of our recommendation system on standard benchmarks: SVM, NB, C4.5 [11] for classification techniques on sentiment analysis tasks. The results were promising. Furthermore, we also experimented with various routing algorithms (A* [10], Yen's [16], and Dijkstra [3]) used to rank the places' distances. We evaluated how they affect the relevance of search results in terms of the F-measure, G-measure [5], and M-measure [1].

## 2  MODEL DESCRIPTION

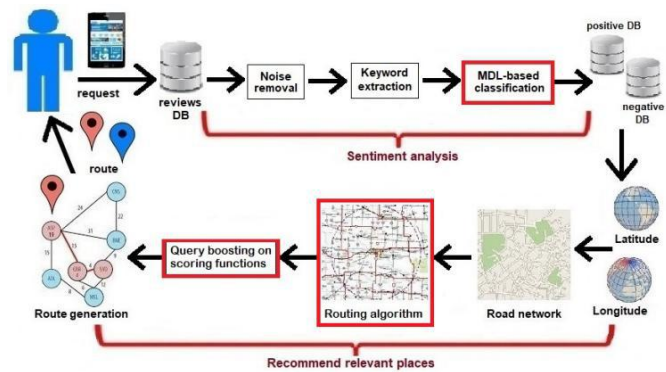

**Figure 1: The proposed highly relevant routing recommendation system model**

Fig. 1 illustrates the schema of our proposed route recommendation system model. We have implemented this model and evaluated its performance later (see Section 3 for experimental setup and Section 4 for the results). There are two main stages in the system: first is the sentiment analysis and second is the recommendation stages. In the first stage, the database of visitor reviews as the knowledge source of the recommendation system is classified into two sentiments: either positive sentiment or negative sentiment. This process of classifying knowledge source can be done offline. The corresponding information retrieval process involves: noise or stopword removal, keyword extraction using RAKE [12], and sentiment classification. In order to do the last part, a classifier must have proper vocabularies of positive, negative, and negation words. In the recommendation stage, the user poses a query to find a place. His/her position (latitude and longitude) is then detected and a road network within some predefined radius is generated. A shortest path routing algorithm is used to compute distances for each candidate place that meets the query's criteria. A scoring function is then used to score the candidates. Finally, the system returns a visualization of the routes to the user.

We highlight some points about the proposed routing recommendation system model: 1) a classification technique that can handle small sized data such as user reviews, 2) query boosting to improve recommendation quality, and 3) a proper selection of the routing algorithm. The first two will be described in the following subsections in this Section 2. The last one will be discussed in the Section 6 on our experiments' results.

### 2.1  Sentiment Analysis

We will use an MDL-based classification to do the sentiment analysis; more specifically the KRIMP algorithm [15]. The algorithm mines frequent patterns of itemsets in a database. Different from Apriori, in KRIMP the frequent itemsets are mined by only picking frequent patterns that compress the database best (see Fig. 2, left).

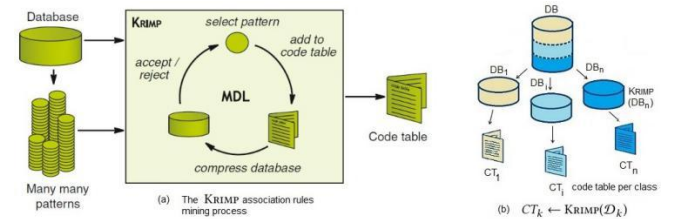

**Figure 2: Preprocessing activities**

All frequent patterns are put in a code table, which in a sense is a representative of the original database, but smaller in size. To use KRIMP as a classifier, a training database $\mathcal{D}$ which has been tagged with the used class labels is needed. Then $\mathcal{D}$ is split to construct a partition $\mathcal{D}_k$ for each class $k$, to derive the corresponding code table $CT_k$, see Fig. 2 (right). All class labels are then removed from all transactions (rows). After some pruning process, all $CT_k$ is incremented by 1 to ensure that their code length can be calculated, thus a transaction can be encoded. For pruning, KRIMP uses post-acceptance pruning. That is, KRIMP only prunes when a candidate pattern $F$ is accepted. $F$ is accepted when the candidate code table $CT_c = CT \cup F$ is better than $CT$, i.e. $L(\mathcal{D},CT_c) < L(\mathcal{D},CT)$.

Each code table $CT_k$ will act as the classifier for $k$. Algorithm 1 shows the classification algorithm. Line 2 constructs all the code tables. For now we should ignore line 3 and pretend that $\mathcal{D}''$ is the database of user reviews. To classify each transaction $t \in \mathcal{D}''$, we first calculate the encoding of $t$ according to all code tables. By the MDL property of KRIMP, $t$ has a higher probability to belong to a class $k$ if the length of its encoding according to $CT_k$ is shorter. So, we can classify $t$ by assigning it to the class whose code table yields the shortest encoding of $t$. This is determined by the loop in lines 7-14.

**Algorithm 1:** Classifying degenerated data with MDL

**Input:** $\mathcal{D}$: database, $t$: transaction, $\delta$: percentage of truncated factor, $class$: class labels in $\mathcal{D}$
**Output:** $\mathcal{D}''_k$: set of degenerated transactions per class

1 **def** $TruncateClassification(\mathcal{D},t,\delta)$:
2      $CT_k \leftarrow \text{KRIMP}(\mathcal{D}_k)$, for all $k \in class$
3      $\mathcal{D}'' \leftarrow truncate(\mathcal{D},\delta)$
4      **forall** $t \in \mathcal{D}''$ **do**
5          $l_{min} \leftarrow \infty$
6          $winner \leftarrow -1$
7          **forall** $k \in class$ **do**
8              **forall** $c \in CT_k$ **do**
9                  $l_{CT_k}(c) = -\log(\frac{freq(c)}{\sum_{d \in CT_k} freq(d)})$
10              $L_{CT_k}(t) = \sum_{c \in cover(t,CT_k)} l_{CT_k}(c)$
11              $l_t \leftarrow L_{CT_k}(t)$
12              **if** $l_t < l_{min}$ **then**
13                  $l_{min} \leftarrow l_t$
14                  $winner \leftarrow k$
15          $\mathcal{D}''_{winner} \leftarrow \mathcal{D}''_{winner} \cup \{t\}$
16      **return** $\mathcal{D}''_k$, for all $k \in class$

More precisely, line 9 calculates the optimal code length of each code in each code table. If $freq(c)$ is the frequency of itemset $c$ in an code table $CT_k$, then we normalize it with the total frequency of all itemsets in the table. Since frequency determines probability of tag co-occurences in the database, the more frequent an itemset reduces the size of the databases the higher probability it has. This is the use of $-log$ in Step 9. Step 10-14 aims to find the shortest code length among the $CT_k$'s. The winner class is the class $k$ whose $CT_k$ yields the shortest code length. Finally $\mathcal{D}''$ is partitioned according to the classification, to yield partitions $\mathcal{D}''_k$ for each class $k$ (line 15).

For the purpose of the experiments in Section 6 we want to simulate a database containing short user reviews. So in line 3 we introduce a truncation step. We create a degenerated database $\mathcal{D}''$ by randomly dropping some tags of every transaction $t$ in $\mathcal{D}$. Only a percentage of (1-$\delta$) of the lexicographically ordered tags of $t$ will be picked; $\delta$ is data degeneration factor or truncated factor. A tag can be a term or a word phrase that represent subset of a document (e.g. visitor reviews database).

## 2.2 The Recommendation Step

During the recommendation step we need to rank/score the resulting candidates that match the given query. As the starting point, we use the scoring function as in [6] as it allows us to give more weight to the more interesting documents and to documents that contain query phrases rather than individual query terms. More precisely, given a user query $q$ and a result document $d$, this scoring function is defined by:

$$score(q,d) = queryNorm(q) * coord(q,d) * \\ \sum_{i=1}^{n} \left( tf(t_i,d) * (idf(t_i))^2 * boost(t_i,d) * norm(t_i,d) \right)(t \in q) \quad (1)$$

where $queryNorm(q)$ is the query normalization used to find the most interesting documents that meet the user query; $coord(q,d)$ is the query coordination used to give more importance on phrases than individual terms. Field-length normalization $norm(t_i,d)$ is used to measure the importance of a term $t_i \in q$ in the document $d$, with regards to their field length. Other scoring elements are: *tf*, *idf*, and boosting factors. To support the goal of building a model for highly relevant recommendation systems that can suggest desired nearest place to the user, we must also assign a robust classification technique to handle few data, we must also take other supporting factors into account such as: the distance and the place's popularity to improve the quality of information.

*Boosting Factors.* To improve the scoring we propose the following boosting function, capturing several key aspects such as the physical routing distance between a candidate place and the user (we would prefer closer candidates) and, importantly, the place's sentiment as inferred from its reviews:

$$boost(t_i,d) = w_{length}(q,d) * w_{sentiment}(t_i,d) * \\ w_{dist}(q,d) * w_{pop}(t_i,d) * w_{field\_matching}(t_i,d) \quad (2)$$

where each boosting component is defined as below. The scores of boosting factors are in the range of 0-3 as the average number of component query terms in searching system is 3 terms [14].

The five proposed boosting factors are as follow:

(1) **Length of Matching Boosting.** $w_{length}(q,d)$ is the weight of the length of matching boosting with its score equal to the number of unique query terms $t_i$, for all $t_i \in q$ found in the document $d$, capped at three.
(2) **Sentiment Boosting.** $w_{sentiment}(t_i,d)$ is the weight of the sentiment of document $d$: it can be upgraded if $d$ has positive sentiment or otherwise downgraded by half of $w_{length}(q,d)$. Since $w_{sentiment}(t_i,d)$ is influenced by the quality of the classification technique that is used on the knowledge database (e.g. on visitors' comments), then transactions (or rows) in the database need to be classified correctly to the class they belong to --Section 2.1 describes the process.

(3) **Distance Boosting.** $w_{dist}(q,d)$ is the weight of the distance boosting, where the score of 3 corresponds to a distance of ≤ 1 km, score=2 to 1 < distance ≤ 2 km, and score=1 to 2 < distance ≤ 5 km. We assume that the first situation represents "walking distance", and the other two represent "driving distance". Finally, the score is set to 0 if the distance > 5 km to indicate "too far away".

(4) **Popularity Boosting.** $w_{pop}(t_i,d)$ is the user's of popularity boosting built under the assumption that popular public places are usually indicated by the user's statement of a place name and/or a street name. Thus if the term $t_i$ matches in both *d.name* and *d.address* than score=3, if $t_i$ only matches in either *d.name* or *d.address* then score=2, if $t_i$ only matches in *d.review* then score=1, and if no match score=0.

(5) **Field Matching Boosting.** $w_{field\_matching}(t_i,d)$ is the weight of field matching boosting with score=3 if $t_i$ is matched in *d.name*, score=2 if $t_i$ is matched in *d.review*, score=1 if $t_i$ is matched in *d.address*, and score=0 if no match.

All in the boosting factors were built under assumption that a typical review consists of place name, address of the place, and what a visitor comment about the place.

## 3 EXPERIMENTAL DESIGN

### 3.1 Data

We use nearest restaurant recommendation as the case study, using the datasets from FourSquare[2] and Yelp Dataset Challenge[3]. In particular, we collected restaurant dataset from FourSquare from 6 February 2016 - 7 April 2016 which results in a text corpus of 2800 words from 699 transactions and restaurant review datasets from Yelp, resulting in a text corpus with more than 267450 words from 7999 transactions. All is compressed in MDL with minimum support equal to 1. It has a very low score since we want to capture more associations but only few corpora exists on many transactions in the databases.

### 3.2 Evaluation

Classification results were measured using several metrics: precision, recall (sensitivity), $F_1$, true negative rate (specificity), accuracy, and predicted positive condition rate (ppcr).

To measure database dissimilarity in classification of sentiment analysis of user reviews, we hold that: given two CTs, e.g. $CT_1$ and $CT_2$ respectively, then $CT_2(t) - CT_1(t)$ measures how characteristic $t$ is for database $\mathcal{D}_1$; see [15].

Search results ranking were measured using F, G, and M measures. F-measure [5] is an extension of Kendall's tau that is used to handle when the compared recommendation systems use routing algorithms that rank non-identical sets of elements (URL of restaurants). F attains the value 1 when the two lists are identically ranked and the value 0 when the lists appear in the opposite order. G-measure works contrarily to F. G altogether with M is used to know if the lists are in high positions. F, G, M all have scores in the range 0-3.

F-measure [5] is defined by extension [1]:

$$F^{k+1}(\tau_1, \tau_2) = 2(k-z)(k+1) + \sum_{i \in Z} |\tau_1(i) - \tau_2(i)| - \sum_{i \in S} \tau_1(i) - \sum_{i \in T} \tau_2(i) \quad (3)$$

where $k$ is length of comparing list, $Z$ is the set of overlapping elements, $z$ is the size of $Z$, $S$ is the set of elements that only appear in the first list of two comparing lists, and $T$ is the set of elements that only appear in the second list. $k+1$ symbolized the extended case in which the two comparing lists are not identical, so by $F^{(k+1)}(\tau_1,\tau_2)$ there is an arbitrary placement larger than the length of the list to documents appearing in one of the lists but not in the other. $\tau_1(i)$ is the Kendall's tau for $S$, and $\tau_2(i)$ is the Kendall's tau for $T$. G-measure [5] is a normalization of $F^{(k+1)}$, over all lists of queries and search routing algorithms, defined by:

$$G^{k+1} = 1 - \frac{F^{(k+1)}}{\max(F^{k+1})} \quad (4)$$

Let:

$$M' = \sum_{i \in Z}\left(\frac{1}{rank_1(i)} - \frac{1}{rank_2(i)}\right) + \sum_{i \in S}\left(\frac{1}{rank_1(i)} - \frac{1}{(len_2+1)}\right) + \sum_{i \in T}\left(\frac{1}{rank_2(i)} - \frac{1}{(len_2+1)}\right) \quad (5)$$

where $rank_1(i)$ is the rank of element $i$ in the first set and $rank_2(i)$ is its rank in the second set. $len_1$ and $len_2$ each is a length of $list_1$ and $list_2$. $S$ is the set of elements that appear in the first list but not in the second list. $T$ is the set of elements that appear in the second list, but not in the first list. $M$ is normalization of $M'$ and defined by $M = 1 - \frac{M'}{4.03975}$ [1]. $M$ has the advantage of handling documents that are not indexed at all by other search routing algorithm.

## 4 RESULTS

Three experiments were conducted. The first two evaluate the performance of MDL-based sentiment analysis. The third experiment investigates the interplay with the used routing algorithm. In Table 1 we used KRIMP algorithm with small minimum support (minsup=3) to compress the database. We set the minsup with small number to have representative frequent pattern item sets of the original database. As Fig. 2 stated, we divide the database into two classes of user reviews' sentiment analysis, the positive class and the negative class. By MDL compresion (KRIMP) we obtain the code table (CT) from each classes that later they are used to measure the distance of a user review (or a "transaction") to each of the classes. The transaction will belong to a class with shorter distance. The typical KRIMP output is a CT consists of non singleton and singleton item sets. The optimal set of frequent item sets is defined as the item sets in which its CT can minimizes the total compressed size of CT and the size of the database ($|\mathcal{D}|+|CT|$). It is clear from Table 1 that the Yelp databases contains more data than the FourSquare database, with maximum length up to 292 tags per transaction,

---
[2] https://developer.foursquare.com/docs/api/venues/explore
[3] http://www.ics.uci.edu/~vpsaini/

whereas in FourSquare it is a very small database with maximum length of 10 tags per transaction. To simulate small data items we apply truncation on the used evaluation databases. This is as explained in Section 2.1 ---see again the explanation of line 3 in Algorithm 1. To simulate increasingly lower information density (in user reviews), we try different truncation/degeneration factors, namely δ= {0%, 25%, 33%, 50%, 67%}.

**Table 1: Database Characteristics**

| Database | Class | #Transactions | #Tags | Max Length | #Candidates | #Used Tags | #Itemsets | $|\mathcal{D}|$ | $|CT|$ | totalSize= $|\mathcal{D}|$+$|CT|$ | min sup |
|---|---|---|---|---|---|---|---|---|---|---|---|
| Yelp_1 | Pos | 345 | 16794 | 260 | 86127 | 12171 | 243 | 257145 | 351637 | 608781 | 3 |
| Yelp_1 | Neg | 55 | 16794 | 233 | 1512 | 2566 | 26 | 36525 | 60021 | 96546 | 3 |
| Yelp_2 | Pos | 330 | 16983 | 292 | 56458 | 11313 | 227 | 238653 | 324230 | 562883 | 3 |
| Yelp_2 | Neg | 70 | 16983 | 184 | 3940 | 3762 | 53 | 58013 | 93125 | 151139 | 3 |
| Yelp_3 | Pos | 350 | 17075 | 269 | 96013 | 12369 | 248 | 258549 | 358035 | 616584 | 3 |
| Yelp_3 | Neg | 50 | 17075 | 164 | 1623 | 2635 | 34 | 37542 | 62114 | 99656 | 3 |
| FourSquare | Pos | 630 | 2800 | 10 | 7089 | 296 | 315 | 5806 | 18588 | 24394 | 1 |
| FourSquare | Neg | 59 | 2800 | 7 | 580 | 26 | 33 | 347 | 1275 | 1622 | 1 |

## 4.1 MDL-based Classification on Large Size Database

The first experiment assumes that we have access to a large amount of user reviews. For this experiment we use the Yelp challenge dataset. Table 2 shows the result, the MDL technique KRIMP is very strong. It outperforms linear and RBF SVM in terms of accuracy, precision, and $F_1$ measure, implying that the code tables produced by KRIMP form a very good model of the original database. Further, we can see that classification performance of KRIMP is mostly influenced by δ: the performance decreases as δ increases. In MDL, a transaction $t$ is assigned a class label belonging to the code table with the minimal encoded length. Therefore the predicted positive condition rate (PPCR) of KRIMP is lower than SVM, since it uses code tables (models), not the total percentage of the total population. This is also the case why MDL-based classification has lower scores than the SVMs on terms of sensitivity and specificity.

If Table 2 shows results where the evaluation sets are the same with the training sets, in Table 3, we further investigate the effect of different coverage testing area, denoted by the cross-validation training sets parameter, given seen and unseen evaluation data on certain percentages of truncation. In Table 3, we divide training and testing datasets separately with 80:20 percentages (seen:unseen percentages) of total number of transactions in the database. Then for training set we take the cross-validation set up (or CV for short) into account. If the CV=1 means that we use all the 80% of total number of transactions in the database as the training dataset, while CV={2,5,10} means we divide the training database (viz. the 80% of total database) into either 2, or 5, or 10 splits respectively and used the CV-1 splits that combined together as the training set whereas the testing set is taken from the remain one split.

From Table 3, it clearly shows that the accuracy of the data will be dropped as the percentage levels of the data degeneration were increased. The situation mostly to occur in unseen testing data. From Table 3, we also observe that although RBF SVM is better than KRIMP on seen testing data but RBF SVM's performance decrease as the truncation factor increase. The Table 3 shows that the RBF SVM is not as stable as KRIMP performance when handling data degeneration. The KRIMP, in other hand, is still maintain very high accuracy on unseen testing data regardless the truncated factors. Fig. 3 gives a closer look of Table 3. From Table 3 and Fig. 3, it shows that the more training instances trained to the system the better the accuracy of the classification.

**Table 2: The Sentiment Analysis Classification Performance on Large Size Database**

| Database | Classifier | Degeneration(%) | Precision (%) | Recall (%) | $F_1$ (%) | TNR (%) | Accuracy (%) | PPCR (%) |
|---|---|---|---|---|---|---|---|---|
| Yelp | KRIMP | 0 | 98.986 | 98.58 | 98.78 | 14.054 | 97.934 | 85.037 |
| Yelp | KRIMP | 25 | 98.949 | 98.05 | 98.5 | 14.082 | 97.446 | 84.613 |
| Yelp | KRIMP | 33 | 98.966 | 96.95 | 97.94 | 14.138 | 96.527 | 83.751 |
| Yelp | KRIMP | 50 | 99.082 | 95.28 | 97.14 | 14.805 | 95.212 | 81.851 |
| Yelp | linear SVM | 0,25,33,50 | 85.4 | 100 | 92.11 | 0 | 85.4 | 100 |
| Yelp | RBF SVM | 0,25,33,50 | 89.295 | 99.98 | 94.25 | 27.56 | 89.415 | 95.94 |

**Table 3: The Sentiment Analysis Classification Performance on Large Size Database with Diverse Testing Area**

| Database | CV | Algorithm | Evaluation Type | δ=0% | δ=25% | δ=33% | δ=50% | δ=67% |
|---|---|---|---|---|---|---|---|---|
| Yelp | 1 | RBF_SVM | Seen | 100 | 100 | 100 | 100 | 100 |
| Yelp | 1 | Linear_SVM | Seen | 85 | 85 | 85 | 85 | 85 |
| Yelp | 1 | KRIMP | Seen | 99 | 100 | 99 | 99 | 99 |
| Yelp | 2 | KRIMP | Seen | 79 | 78 | 77 | 72 | 52 |
| Yelp | 5 | KRIMP | Seen | 80 | 77 | 77 | 75 | 58 |
| Yelp | 10 | KRIMP | Seen | 80 | 77 | 77 | 72 | 59 |
| Yelp | 1 | RBF_SVM | Unseen | 100 | 89 | 87 | 87 | 85 |
| Yelp | 1 | Linear_SVM | Unseen | 85 | 85 | 85 | 85 | 85 |
| Yelp | 1 | KRIMP | Unseen | 100 | 100 | 100 | 100 | 100 |
| Yelp | 2 | KRIMP | Unseen | 77 | 68 | 65 | 52 | 40 |
| Yelp | 5 | KRIMP | Unseen | 74 | 68 | 73 | 56 | 43 |
| Yelp | 10 | KRIMP | Unseen | 76 | 70 | 74 | 58 | 45 |

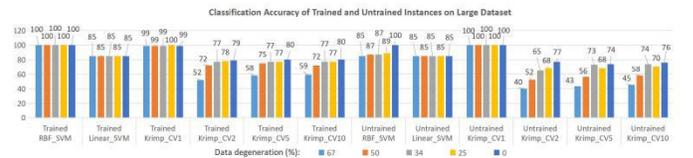

**Figure 3: Classification performance on large database**

Fig. 4 is a sample of how the size of the encoded database in total influences the accuracy score. From Fig. 4 even for the very few tags database case, such as the FourSquare database with maximum length of 10 tags per transaction that truncated into only 33% of data remaining, we can see that the larger coverage size of the original database it has, denoted by its Total Encoded Size, the better the average score the MDL model has. Or one can conclude that the large the size that is covered by the encoded database, the closer the MDL model to the original database thus maintaining the good accuracy of the MDL-based classification system.

Fig. 5 is an example of CT in a small database and its predicted class given a transaction. It clearly shows that different set up of coverage area on training dataset will create different CT scope that impact on the code length difference (see $L_{CTk}(t)$ in Algorithm 1). The inexistence of some tags on the training dataset impact on their absence on CT; this kind of situation potential to lead to such failure in the MDL-based classification (see CTs in the middle of Fig. 5 that represents a good example of query of set-of-rare tags).

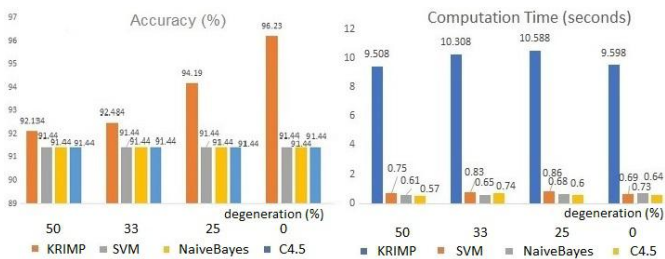

**Figure 5: An example of CT and class in FourSquare database while given different coverage area of training set**

## 4.2 MDL-based Classification on Small Size Database

In the second experiment we evaluate the performance of KRIMP on a smaller database, to simulate the situation where we do not yet have much user reviews as our base. For this, we use the FourSquare datasets. Table 4 shows the result, while Fig. 6 provides a visualization derived from Table 4. The results clearly show that in the first experiments kind, MDL based classification significantly outperforms other classification techniques. Following are comparisons of the total running time to do sentiment analysis tasks: KRIMP:SVM=13:1, KRIMP:Naive Bayes=15:1, and KRIMP:C4.5=16:1. KRIMP is the slowest one but still slightly fast (the worst case based on Table 4 is below 11 seconds).

**Table 4: The Sentiment Analysis Classification Performance on Small Size Database**

| Database | Degeneration (%) | Accuracy (%) | | | | Time (s) | | | |
|---|---|---|---|---|---|---|---|---|---|
| | | KRIMP | SVM | NB | C4.5 | $t_{KRIMP}$ | $t_{SVM}$ | $t_{NB}$ | $t_{C4.5}$ |
| fourSquare | 0 | 96.23 | 91.44 | 91.44 | 91.44 | 9.598 | 0.69 | 0.73 | 0.64 |
| fourSquare | 25 | 94.19 | 91.44 | 91.44 | 91.44 | 10.588 | 0.86 | 0.68 | 0.6 |
| fourSquare | 33 | 92.484 | 91.44 | 91.44 | 91.44 | 10.308 | 0.83 | 0.65 | 0.74 |
| fourSquare | 50 | 92.134 | 91.44 | 91.44 | 91.44 | 9.508 | 0.75 | 0.61 | 0.57 |

**Figure 6: Classification performance on small database**

In general, from Table 2, Table 3 and Table 4 show that until 50% of data degeneration, well-known classifiers such as SVM, Naive Bayes, and C4.5 can maintain their performance. Although the classifiers show good results but their performances were still below the MDL-based (viz. KRIMP) accuracy, both in large size database or small database.

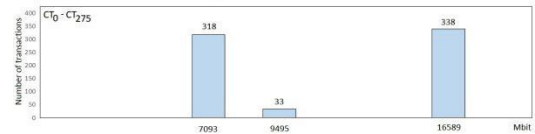

**Figure 7: An example of histogram of code length difference for transaction in FourSquare database ($CT_0$ - $CT_{275}$)**

Fig. 7 is the code length difference histogram for encoded length difference between $CT_0$ and $CT_{275}$. From Fig. 7, the distribution of code length is concentrated only on certain large size of bits that indicates that the two user reviews' sentiment classes, both positive class and negative class, are easy to classify due to they are highly distinct in characteristics. If the code length difference is in small bits range, it means the classes have more similarity in characteristics.

## 4.3 Improving The System Performance Using Embedded Boosting Factors

In the final experiment, we investigate the influence of incorporating reviews' sentiment to the resulting recommendation. As comparison, we implemented three routing algorithms: Dijkstra [3], A* [10], and Yen [16]. They may produce different routes (and therefore also different distance). We test each one of them, with the *boosting factors* described in Section 2.2 *turned on* (hence it will take sentiment into account), and compare the resulting recommendation list with that of using Dijkstra *without* the boosting factors (hence ignoring other boosting factors but the distance factor). As the database we use the FourSquare datasets. There are 90 queries used in the experiments. We use a different natural formulation of user queries, for example a query could mention either the name or address of the inquired restaurant, instead of just mentioning food names. The motivation is to test if the retrieval system can retrieve relevant recommendations as user expects, viz. by considering popularity and nearness of a place, field and query matching, as well as that the recommended place must originated from a place that categorized has positive sentiment.

Fig. 8 and Fig. 9 show the results, with the latter showing more details. The *x*-axis represent the number of overlapping elements (normalized to [0..1.0]); we divide that in five equal length intervals of size 0.2 in decreasing order. E.g. *x*=1 represents the interval [0.8..1], *x*=2 the interval [0.6..0.8) etc. We compare the routing algorithms in pairs. The boosting factors are turned on, except on one variant of the Dijkstra algorithm denoted by Dijkstra$_{norel}$, where the boosting factors are turned off. The compared pairs are: the pair A compares (A*,Dijkstra$_{norel}$), the pair D compares (Dijkstra,Dijkstra$_{norel}$), and E compares (Yen,

Dijkstra$_{norel}$). Additionally, the pair B compares (A*,Dijkstra), C = (A*,Yen), and F = (Yen,Dijkstra).

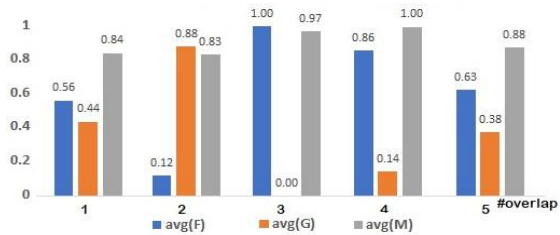

**Figure 8: System performance on overlapped URLs**

Fig. 8 shows that the *M*-measure is almost always considerably higher than the *G*-measure. It means that, in general, the overlapping elements in search results are in highly ranked. However, in the overlapping range of [0.6..0.8) *G* is actually higher than *M*, implying that the overlapping elements tend to appear at the opposite sides of the recommended lists.

Fig. 9 suggests that the best routing algorithm for the recommendation model is either A* or Yen. The use of Dijkstra$_{norel}$ is not suggested because its relevant search results are in the lower positions. Furthermore, the trend of G movements is the opposite to F, while F movements are similar to M. By considering its formula, the M-measure is good in handling overlapping areas and in handling comparisons of lists with different length. With M ≥ 0.71 in average, and with the G scores mostly lower than the M scores, as well as the high average of the F scores (F=0.72), these indicate that the proposed routing recommendation system model is very strong both in returning highly relevant search results on top positions as well as in returning most recommended (overlapped) search results. Its total running time is acceptable (only 11.25 seconds).

## 5 CONCLUSIONS

In this work, we have focused on providing high quality recommendation system model using boosting factors in scoring function and the use of MDL principles on classification tasks. We have shown that using MDL-based classification we can handle classification on small data items very well. Our findings indicate that improvements in classification can be achieved by using code table from large datasets. We have shown that coverage area in training dataset has strong impact on classification accuracy. The larger the size that is covered by the encoded database the closer its MDL model to the original database thus will result in very good accuracy. For the few or limited database, although the performance of the MDL-based classification degrades on when the data items grow shorter, simulated by increasing degeneration factor δ, it still outperforms other state of the art classification techniques --we permit up to 67% of data loss. Next, we have found that a proper selection of shortest routing algorithm (e.g. A* or Yen's) can provide further gains in quality. Finally, adding boosting factors can provide further gains in quality of the recommendation system model. The recommendation system model described in this paper can be implemented in various recommendation projects that use routing data and face the small data instances problem on its knowledge source.

## ACKNOWLEDGMENTS

The authors would like to thank: Y. Pinata, Julio Fernando, Edo Afriando, and Rina Rahmadini,  for preparing the dataset.

The authors would also like to thank the anonymous referees for  their valuable comments and helpful suggestions.

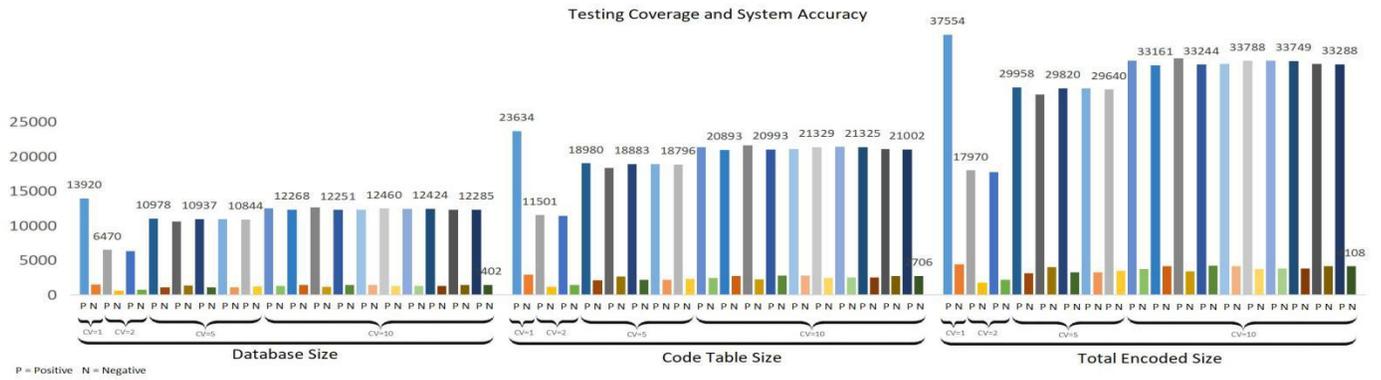

**Figure 4: Testing coverage vs system performance.** In this example the testing data for the Yelp_3 with data degeneration of 67% has variety range of accuracy (in average) depend on the coverage area viz. the cross validation testing area. For CV = {1, 2, 5, 10} the average accuracy is {0.9925, 0.53, 0.575, 0.5825} respectively.

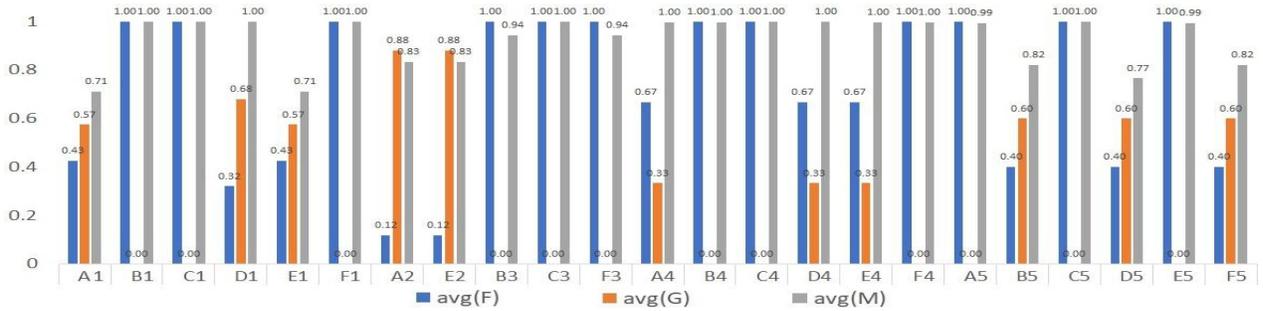

**Figure 9: System performance on routing algorithm**